\documentclass[article,11pt,onecolumn,superscriptaddress,showpacs] {revtex4-1}
\usepackage{amsmath}
\usepackage{latexsym}
\usepackage{amssymb}
\usepackage{graphicx}
\usepackage[colorlinks=true, citecolor=blue, urlcolor=blue]{hyperref}
\usepackage{float}
\usepackage{amsfonts}
\begin{document}
\title{Several foundational and information theoretic implications of Bell's theorem}

\author{Guruprasad Kar}
\email{kar.guruprasad@gmail.com}
\affiliation{Physics and Applied Mathematics Unit, Indian Statistical Institute, 203 B.T. Road, Kolkata-700108, India}

\author{Manik Banik}
\email{manik11ju@gmail.com}
\affiliation{Physics and Applied Mathematics Unit, Indian Statistical Institute, 203 B.T. Road, Kolkata-700108, India}

\begin{abstract}
In 1935, Albert Einstein and two colleagues, Boris Podolsky and Nathan Rosen (EPR) developed a thought experiment to demonstrate what they felt was a lack of completeness in quantum mechanics. EPR also postulated the existence of more fundamental theory where physical reality of any system would be completely describe by the variables/states of that fundamental theory. This variable is commonly called hidden variable and the theory is called hidden variable theory (HVT). In 1964, John Bell proposed an empirically verifiable criterion to test for the existence of these HVTs. He derived an inequality, which must be satisfied by any theory that fulfill the conditions of \emph{locality} and \emph{reality}. He also showed that quantum mechanics, as it violates this inequality, is incompatible with any \emph{local-realistic} theory. Later it has been shown that Bell's inequality can be derived from different set of assumptions and it also find applications in useful information theoretic protocols. In this review we will discuss various foundational as well as information theoretic implications of Bell's inequality. We will also discuss about some restricted nonlocal feature of quantum nonlocality and elaborate the role of Uncertainty principle and Complementarity principle in explaining this feature.    
\end{abstract}
\maketitle
%\keywords{Quantum Mechanics; Bell's inequality; Nonlocality; Complementarity}

\section{Introduction}

Quantum mechanics (QM) is, certainly, the most successful theory to describe physical phenomena at very small scales. Apart from gravity, it provides completely correct mathematical description for all natural phenomena, stating from the structure of atoms, the rules of chemistry and properties of condensed matter to nuclear structure and the physics of elementary particles. However, from a fundamental point of view different concepts of this theory departs in various ways from that of classical physics. Undoubtedly the most debatable \emph{non classical} concept is the existence of \emph{nonlocal} correlations among spatially separated quantum systems. John Bell, a Northern Irish physicist, in 1964, established a seminal result, that certain quantum correlations, unlike all other correlations in the Universe, cannot arise from any local cause \cite{Bell,Gottfried}. This result is commonly known as `Bell's theorem'. For the last $50$ years, this theorem remains at the center of almost all \emph{metaphysical} debates of Quantum foundation and during the last two decades it also finds applications in various information theoretic tasks.   

In his paper Bell addressed a long standing question introduced by Einstein along with his two colleagues Podolsky and Rosen (EPR) at the very early days of quantum theory \cite{EPR}. From historical perspective, Einstein was the most prominent opponent of the `Copenhagen interpretation', according to which quantum system is completely described by a wave function belonging to an Hilbert space associated with the system \cite{Copenhagen}. He believed that the fundamental theory of the nature must be deterministic. It was the intrinsic randomness of QM (lack of reality) which bothered him the most. His dissatisfaction with QM, in describing the natural phenomena, comes clear from one of his famous comment ``\emph{God does not play dice with the universe}''. In his view, though QM provide the correct description of physical phenomena it could not be the complete one. In the groundbreaking $1935$ EPR paper this philosophical idea has been condensed into a physical argument. In this paper the authors first described a necessary condition of completeness for a physical theory:
\begin{enumerate}
\item[] {\bf Condition of completeness}: \emph{Every element of the physical reality must have a counter part in the physical theory}.
\end{enumerate}    
Once the the condition of completeness is given, naturally the question arises: `what are the elements of the physical reality?' EPR did not provide a comprehensive definition of reality, but, they described a sufficient condition of it:  
\begin{enumerate}
\item[] {\bf Physical reality}: \emph{If, without in any way disturbing a system, we can predict with certainty (i.e., with probability equal to unity) the value of a physical quantity, then there exists an element of physical reality corresponding lo this physical quantity.}
\end{enumerate}
Invoking, then, the well established idea from special theory of relativity, that nothing can travel faster than light, EPR succeed to demonstrate that ``\emph{wave function does not provide a complete description of the physical reality}''. EPR ended there paper in the following fashion ``\emph{....we left open the question of whether or not such a description exists. We believe, however, that such a theory is possible}''. EPR work immediately gave rise to intense debates among physicists as well as philosophers about the foundational status of QM. Within few days, an important critique appeared by N. Bohr, a pioneer proponent of the Copenhagen interpretation, against the EPR argument \cite{Bohr}. But, for a long period of time the debate was only a subject of theoretical interest, no empirical testable method was known to resolve this debate. John Bell took the challenge whether some experimental criteria can be derived to address this debate and ultimately in $1964$ he succeeded to derive it as an powerful theorem. This theorem tells that there exist quantum correlations which are incompatible with conjunction of two classical notion, namely, \emph{locality} and \emph{reality}. However, later it has been shown that the same Bell's theorem can be derived under different sets of assumptions. Violation of this theorem implies different consequences depending on under which set of assumptions the theorem is derived.

This paper is organized in the following manner. In Section (\ref{sec2}) we first discuss operational and ontological framework for any physical theory. We then discuss the Bell's original derivation of his famous inequality under the assumptions of \emph{locality} and \emph{reality}. We also discuss about the quantum mechanical violation of this inequality. In Section (\ref{sec3}) we discuss that Bell's inequality (BI) can be derived under the assumption of joint measurability and relativistic causality or no signaling principle. In Section (\ref{sec4}) we analyze that the same BI can be derived under some operational assumptions. We also discuss that as a novel implication of this derivation BI finds application in an important information theoretic task, namely device independent randomness certification. In Section (\ref{sec5}) we discuss another important feature of quantum correlation namely the limited violation of BI in QM compared to the violation allowed by no-signaling principle. We discuss the role of uncertainty and complementarity principle in explaining this feature.    

\section{Bell's inequality}\label{sec2}
Bell realized that to understand the source of conflicts in QM, it is essential to know where and how our classical concepts and intuitions start to fail in describing the quantum phenomena. He succeeded to derive an experimentally testable condition obtained under some \emph{metaphysical} assumptions which follow from the classical world view. For the first time, in the history of science, there was a concrete suggestion to test the correctness of different world view from measurement results. This area of study is called `\emph{experimental metaphysics}', a term coined by Abner Shimony \cite{Shimony}. Before discussing various assumptions under which BI is derived, we briefly describe the \emph{operational} and \emph{ontological} framework for any physical theory.  

\subsection{Operational interpretation of a theory}
In an operational interpretation of a physical theory, the primitive elements are preparation procedures, transformation procedures, and measurement procedures, which can be viewed as lists of instructions to be implemented in the laboratory. The goal  of an operational theory is merely to specify the probabilities $p(k|M,P,T)$ of different outcomes $k\in\mathcal{K}_M$ that may result from a measurement procedure $M\in\mathcal{M}$ given a particular preparation procedure $P\in\mathcal{P}$, and a particular transformation procedure $T\in\mathcal{T}$; where $\mathcal{M}$, $\mathcal{P}$ and $\mathcal{T}$ respectively denote the sets of measurement procedures, preparation procedures and transformation procedures; $\mathcal{K}_M$ denotes the set of measurement results for the measurement M. When there is no transformation procedure, we simply have $p(k|M,P)$. The only restrictions on $\{p(k|M,P)\}_{k\in\mathcal{K}_M}$ is that all of them are non negative and $\sum_{k\in\mathcal{K}_M}p(k|M,P)=1$ $\forall$ M, P.  As an example, in an operational formulation of quantum theory, every preparation P is associated with a density operator $\rho$ on Hilbert space, and every measurement M is associated with a positive operator valued measure (POVM) $\{E_k|~E_k\ge0~\forall~k~\mbox{and}~\sum_kE_k=I\}$. The probability of obtaining outcome $k$ is given by the generalized Born rule, $p(k|M,P)=\mbox{Tr}(E_k\rho)$. An operational theory does not tell anything about \emph{physical state} of the system. So one can construct an ontological model of an operational theory (for details of this framework, we refer to \cite{Spek05,Rudolph}).

\subsection{Ontological model for a theory}
In an ontological model of an operational theory the primitives of description are the actual state of affairs of the system. A preparation procedure is assumed to prepare a system with certain properties and a measurement procedure is assumed to reveal something about those properties. A complete specification of the properties of a system is referred to as the ontic state of that system. The knowledge of the ontic state may remain unknown to the observer. For this reason it is also called hidden variable and the model is called hidden variable theory (HVT).

In an ontological model for quantum theory, a particular preparation method $P_{\psi}$ which prepares the quantum state $|\psi\rangle$, actually puts the system into some ontic state $\lambda\in\Lambda$, $\Lambda$ denotes the ontic state space. An observer who knows the preparation $P_{\psi}$ may nonetheless have incomplete knowledge of $\lambda$. Thus, in general, an ontological model associates a probability distribution $\mu(\lambda|P_{\psi})$ with preparation $P_{\psi}$ of $|\psi\rangle$. $\mu(\lambda| P_{\psi})$ is called the \emph{epistemic state} as it encodes observer's \emph{epistemic ignorance} about the state of the system. It must satisfy
\begin{equation}\label{epis}
\int_{\Lambda}\mu(\lambda|P_{\psi})d\lambda=1~~
\forall~|\psi\rangle~\mbox{and}~P_{\psi}.
\end{equation}
Similarly, the model may be such that the ontic state $\lambda$ determines only the probability $\xi(k|\lambda,M)$, of different outcomes $k$ for the measurement method $M$. However, when the model is deterministic one, $\xi(k|\lambda,M)\in \{0,1\}$. The response functions $\xi(k|\lambda,M)\in[0,1]$, should satisfy
\begin{equation}\label{response}
\sum_{k\in\mathcal{K}_M}\xi(k|\lambda,M)=1~~\forall~~\lambda,~~M.
\end{equation}
As the model is required to reproduce the observed frequencies (quantum predictions) hence the following must also be satisfied
\begin{equation}\label{reproduce}
\int_{\Lambda} \xi(\phi|M,\lambda)\mu(\lambda|P_{\psi}) d\lambda = |\langle\phi|\psi\rangle|^2.
\end{equation}     
This requirement is called `quantum reproducibility condition'.

\subsection{Bell's locality-reality theorem}
It is a statement about certain type of correlations established between two spatially separated observers, say Alice and Bob. Each of the observers hold some physical system which may have interacted previously. Denote the preparation state of the composite system as $P$. Both of them perform measurement on there respective subsystem and observe the measurement result. Let, Alice can choose to perform any of two possible measurements $x\in\{0,1\}\equiv X$. Also consider that for each measurement there are two possible outcomes denoted as $a\in\{+1,-1\}\equiv A$. Similarly, Bob measurements are denoted as $y\in\{0,1\}\equiv Y$ and the outcome as $b\in\{+1,-1\}\equiv B$. Now for any pair of Alice's and Bob's measurement settings the expectation value is calculated as
\begin{equation}\label{expect}
\langle xy\rangle_P=\sum_{a,b=+1}^{-1}ab~p(a,b|x,y,P),
\end{equation}
where the subindex $\langle *\rangle_P$ denote that expectation is calculated on the preparation state $P\in\mathcal{P}$. $\{p(a,b|x,y,P)\}_{x\in X,y\in Y}^{a\in A, b\in B}$ are the observed frequency in the operational theory, where $p(a,b|x,y,P)$ denote the joint conditional probability of obtaining outcome `$a$' by Alice and outcome `$b$' by Bob when they perform measurements `$x$' and `$y$' respectively. For this operational experiment one can consider a ontological model as described previously. Denoting the ontological variable (hidden variable) as $\lambda$, the conditional joint probability reads as $P(a,b|x,y,P,\lambda)$. To satisfy the reproducibility condition (\ref{reproduce}) we have  
\begin{equation}\label{reproduce1}
p(a,b|x,y,P)=\int_{\lambda\in\Lambda}d\lambda\rho(\lambda)p(a,b|x,y,P,\lambda).
\end{equation}
The assumptions under which Bell derived his result state certain properties of the ontological theory.

\emph{\bf Definition 1}: An ontological model is said to satisfy locality iff
\begin{eqnarray}
p(a|x,y,P,\lambda)=p(a|x,P,\lambda);~\forall~a,x,y\nonumber\\
p(b|x,y,P,\lambda)=p(b|y,P,\lambda);~\forall~b,x,y.
\end{eqnarray}\label{locality}
 
\emph{\bf Definition 2}: An ontological model is said to deterministic iff it satisfy determinism iff
\begin{equation}\label{determinism}
p(a,b|x,y,P,\lambda)\in\{0,1\};~\forall~a,b,x,y.
\end{equation} 
This implies that $a=a(x,y,P,\lambda)$ and $b=b(x,y,P,\lambda)$.
It is straightforward to check that if any model satisfy locality and determinism then it also satisfy the following factorisability relation
\begin{equation}\label{facto}
p(a,b|x,y,P,\lambda)=p(a|x,P,\lambda)P(b|y,P,\lambda).
\end{equation}

\textbf{ Theorem 1} (Bell): Any theory which satisfy the factorisability condition or in other words satisfy locality and determinism must satisfy the following inequality:
\begin{equation}\label{BI}
|\langle x_0y_0\rangle+\langle x_0y_1\rangle+\langle x_1y_0\rangle-\langle x_1y_1\rangle|\le 2.
\end{equation}
Proof of the above theorem is straightforward and we omit the proof (for the proof we refer the papers \cite{Bell,CHSH}). It is important to note that another implicit assumption, namely the assumption of \emph{measurement independence}, has been used in the derivation of Bell's inequality (BI) \cite{Freewill}. The assumption of measurement independence contravenes to depend the distribution of the ontic variable on the Alice's and Bob's measurement settings, i.e., $\rho(\lambda|x,y)=\rho(\lambda)$. In other word, applying Bay's theorem, we can say that Alice's (Bob's) choices of measurement setting is independent of the ontic variables, i.e., Alice and Bob can choose their measurement setting freely.

Interestingly, quantum statistics violates this inequality. In the following we discuss regarding the violation of BI in QM.

\subsection{Violation of BI by quantum correlation}
Consider two spin-half particle prepared in the state $|\psi^-\rangle_{12}=\frac{1}{\sqrt{2}}(|0\rangle_1\otimes|1\rangle_2-|1\rangle_1\otimes|2\rangle_2)$, where sub-indices have been used to denote first and second particle respectively, and $|0\rangle$ ($|1\rangle$) denotes the up (down) eigenstate of the Pauli $\sigma_z$ operator. The state $|\psi^-\rangle_{12}$ is called singlet state and it belongs to the tensor product Hilbert space $\mathbb{C}^2\otimes\mathbb{C}^2$. This state has an interesting property that it can not be written as mixer of product states of the composite system and hence it is entangled \cite{Werner,Horodecki}. If Alice perform spin measurement along $\hat{n}$ direction and Bob perform along $\hat{m}$ direction on their respective particle of singlet state then according to QM the joint expectation reads as:         
\begin{equation}\label{singlet}
\langle\hat{n}\hat{m}\rangle=_{12}\langle\psi^-|\hat{n}.\vec{\sigma}\otimes\hat{m}.\vec{\sigma}|\psi^-\rangle_{12}=-\hat{n}.\hat{m}=-\cos(\theta_{nm}).
\end{equation} 
Let Alice and Bob fix there measurement setting as following: $\cos(\theta_{n_0m_0})=\cos(\theta_{n_0m_1})=\cos(\theta_{n_1m_0})=\frac{\pi}{4}$ and $\cos(\theta_{n_1m_1})=\frac{3\pi}{4}$. With these measurement settings on singlet sate the Bell inequality expression (i.e. left hand side of Eq.(\ref{BI})) turns out to be $2\sqrt{2}>2$ (loophole free experimental verification of nonlocal nature of QM is an active area of research till date \cite{experiment}). 

It is interesting question to ask what conclusion one can draw from BI violation. As in the previous section we have discussed that BI is derived under the assumption of locality and determinism, so when a theory violates BI we conclude that the description of the theory can not be replaced by an local deterministic ontological model. Since QM violates BI, it can not have a local deterministic description and hence it is called \emph{nonlocal}. Note that to show the violation of BI in QM the measurements that have been chosen on Alice's (Bob's) side are incompatible \textit{i.e} not jointly measurable. In the following section we discuss the connection between incompatibility of observables and BI violation. 

\section{Joint measurability and nonlocality}\label{sec3}
To show the violation of BI in QM we have used two feature of quantum theory, i.e., we have considered (i) entangled system and (ii) incompatible measurement. Naturally the question arise whether these two feature are generic requirement for BI violation in QM. It has been shown that by performing arbitrary local measurements on a separable state it is not possible to establish any form of quantum nonlocality \cite{Werner}. On the other hand, if we want to address the question of incompatible observable, we have to put more general the notion of incompatibility. There are several notions of incompatibility. Two such notions are non-commutativity and non existence of joint measurement \cite{Heinosaari,Busch}. For projective measurement (PVM) these two notions are identical. Andersson \emph{et al.} have shown that BI can be derived under the assumption of existence of joint measurement and no signaling or signal locality condition \cite{Andersson}. It is important to note that if joint measurement exists on both side then the four probability distribution exists and the BI follows according to Fine's result \cite{Fine}. In the following we briefly sketch the proof of Andersson \emph{et al.}  

Consider two measurements $A_1$ and $A_2$ on Alice side and two measurements $B_1$ and $B_2$ on Bob's side. Denote $p[v(A_1)=v(A_2);B]$ be the probability that  $A_1$ and $A_2$ have same measurement outcome when Bob's measurement is $B$. We have
$$p[v(A_1)=v(A_2);B]=p[v(A_1)=v(A_2)=v(B)]+p[v(A_1)=v(A_2)=-v(B)].$$
Positivity condition of the probability implies,
$$p[v(A_1)=v(A_2)=v(B)]+p[v(A_1)=v(A_2)=-v(B)]$$
$$~~~~~~~~~~~~~~~~~~~~~~~~~\ge|p[v(A_1)=v(A_2)=v(B)]-p[v(A_1)=v(A_2)=-v(B)]|.$$
Writing the expectation value as
$$E(A,B)=p[v(A)=v(B)]-p[v(A)=-v(B)]$$
And putting $B=B_1$ we have 
$$p[v(A_1)=v(A_2);B_1]\ge\frac{1}{2}|E(A_1,B_1)+E(A_1,B_2)|$$
Similar argument gives 
$$p[v(A_1)=-v(A_2);B_2]\ge\frac{1}{2}|E(A_1,B_2)-E(A_2,B_2)|$$
Adding the above two inequalities we have 
$$p[v(A_1)=v(A_2);B_1]+p[v(A_1)=-v(A_2);B_2]$$
$$~~~~~~~~~~~~~~~~~~~\ge\frac{1}{2}|E(A_1,B_1)+E(A_1,B_2)|+|E(A_1,B_2)-E(A_2,B_2)|.$$
The no signaling condition implies
$$p[v(A_1)=-v(A_2);B_2]=p[v(A_1)=-v(A_2);B_1].$$
using the normalization of probability,  
$$p[v(A_1)=v(A_2);B_1]+p[v(A_1)=-v(A_2);B_1]=1$$
finally we get 
$$|E(A_1,B_1)+E(A_1,B_2)|+|E(A_1,B_2)-E(A_2,B_2)|\le 2.$$

However, for positive-operator-valued-measures (POVMs) non existence of joint measurability and non-commutativity are not equivalent notions. From now on incompatibility would mean the impossibility of joint measurement. So, at this point one can asks the following two questions: (1) do all entangled states demonstrate nonlocality? (2) Can all pair of incompatible measurements on both sides exhibit nonlocality? It has been shown that though all pure entanglement states violates BI \cite{Gisin}, there exist mixed entangled states which have LHV model for all PVMs \cite{Werner} and also for all POVMs \cite{Barrett}, which means there exist local entangled states or in other words, nonlocality and entanglement are two different concepts. If two incompatible measurements are considered, each with binary outcome, then Wolf et al. have shown that it can always lead to violation of the Bell-Clauser-Horne-Shimony-Holt (Bell-CHSH) \cite{CHSH} inequality \cite{Wolf}. This result has established a connection between (im)possibility of joint measurability and nonlocality.
However, for general situation i.e., for arbitrary number of POVMs with arbitrarily many outcomes the relation between joint measurability and nonlocality is not yet established. The question is important as pairwise joint measurability, in general, does not imply full joint measurability. We discuss two different such examples.

(a) \emph{\bf Orthogonal spin measurement}: Consider the set of three dichotomic POVMs, acting on $\mathbb{C}^2$, given by the following positive operators 
\begin{equation}\label{ortho}
M_{0|j}(\eta)=\frac{1}{2}(\mathbf{1}+\eta \sigma_j)
\end{equation}  
for $j = x, y, z$, where $\sigma_x$, $\sigma_y$, $\sigma_z$ are the Pauli matrices, and $0 \le \eta \le 1$. The operator corresponding to other outcome is $M_{1|j}(\eta)=\mathbf{1}-M_{0|j}(\eta)$. The triple of measurements are pairwise jointly measurable iff $\eta\le\frac{1}{\sqrt{2}}\approx 0.707$, but triple-wise jointly measurable iff $\eta\le\frac{1}{\sqrt{3}}\approx 0.577$.  Hence in the range $\frac{1}{\sqrt{3}}\le\eta\le\frac{1}{\sqrt{2}}$ the the set $\{M_{a|j}(\eta)\}$ forms a hollow triangle \cite{Heinosaari,Liang}.

(b) \emph{\bf Trine spin measurement}: Consider another three dichotomic POVMs, acting on $\mathbb{C}^2$, given by   
\begin{equation}\label{trin}
M_{0|j}(\eta)=\frac{1}{2}(\mathbf{1}+\eta \hat{n}_j.\vec{\sigma}),
\end{equation} 
where, the three vectors $\{\hat{n}_j|j=1,2,3\}$ are equally separated in a plane, i.e. separated by a trine or an angle of $2\pi/3$. This triple of measurements are pairwise jointly measurable if $\eta\le \sqrt{3}-1\approx0.73205$, but triple-wise jointly measurable only if $\eta\le\frac{2}{3}$ \cite{Liang}.

The above two example shows that the connection between joint measurability and nonlocality for more than two measurement scenario, even with each measurement being dichotomic,  is not a trivial extension of that of two measurement scenario. In two recent works partial but very useful progress has been obtained \cite{Quintino,Uola}. The authors in \cite{Quintino} showed that (non) joint measurability can be viewed as equivalent to EPR steering, a strictly weaker type of nonlocality than Bell's nonlocality \cite{Schrodinger,Wiseman}. Specifically, the authors in \cite{Quintino} have shown that for any set of POVMs that are incompatible (i.e. not jointly measurable), one can find an entangled state, such that the resulting statistics violates a steering inequality. They further conjectured that (im)possibility of joint measurability and Bell nonlocality are inequivalent. A conclusive answer requires proof (or disproof) of this conjecture. 

Other than these foundational implications, in the last decade various device-independent \cite{Scarani} information theoretic protocols like cryptography \cite{cry1}, randomness certification \cite{rand1,rand2}, Hilbert Spaces dimension witness \cite{Brunner} have made use of nonlocality arguments \cite{nonlocality}. In the following we discuss one of such protocol, namely, device independent randomness certification. Before that we discuss another derivation of BI based on some \emph{operational} assumptions.

\section{Bell's inequality from operational assumptions}\label{sec4}  
In the previous section we have seen that BI can be derived under the conjunction of two assumption, namely, \emph{locality} and \emph{determinism}. Both these assumptions assert some \emph{ontological} properties. As QM violates BI, it can not have a local-deterministic ontological description. However, there are models that violate locality but maintain determinism (Bohmian mechanics \cite{Bohm} is an example), and models that maintain locality but violate determinism (standard operational quantum theory is an example). Recently, Cavalcanti and Wiseman have derived BI from another set of assumptions, namely, \emph{signal locality} and \emph{predictability} \cite{Cavalcanti}. Unlike the Bell's original assumptions (i.e. locality and determinism), these assumptions are purely operational, that is, they refer to operational quantities only. Whereas the assumption of signal locality describe the impossibility to send signals faster than light, predictability assumes that one can predict the outcomes of all possible measurements to be performed on a system.  

To derive BI from the conjunction of the assumptions signal locality and predictability it is sufficient to show that the joint probabilities of experimental outcomes given by a model is factorisable, i.e., satisfy the Eq.(\ref{facto}). 

\emph{\bf Definition 3}: A model is said to satisfy signal locality iff
\begin{eqnarray}\label{s-locality}
p(a|x,y,P)=p(a|x,P);~\forall~a,x,y\nonumber\\
p(b|x,y,P)=p(b|y,P);~\forall~b,x,y.
\end{eqnarray}

\emph{\bf Definition 4}: A model is said to be predictable, or to satisfy predictability iff
\begin{equation}\label{predict}
p(a,b|x,y,P)\in\{0,1\};~\forall~a,b,x,y.
\end{equation}
This implies that $a=a(x,y,P)$ and $b=b(x,y,P)$.

\textbf{Lemma 1} (Cavalcanti \& Wiseman): signal locality $\wedge$ predictability $\Rightarrow$ BI.

\emph{Proof}: The definition of predictability implies that
$$p(a,b|x,y,P,\lambda)=p(a,b|x,y,P)$$
as $p(a,b|x,y,P)\in\{0,1\}$, conditioning on $\lambda$ can not alter it. Using Bay's rule of conditional probability
$$ p(a,b|x,y,P)=p(a|b,x,y,P)p(b|x,y,P)$$
Predictability also implies that $p(a|b,x,y,P)=p(a|x,y,P)$. Hence 
$$p(a,b|x,y,P,\lambda)=p(a|x,y,P)p(b|x,y,P)$$
Applying the signal locality condition
$$p(a,b|x,y,P,\lambda)=p(a|x,P)p(b|y,P)$$
Conditioning on the ontic variable $\lambda$, we reached at the desired factorisable condition given by  $$p(a,b|x,y,P,\lambda)=p(a|x,P,\lambda)p(b|y,P,\lambda)$$ 

The above derivation gives authority in drawing remarkable conclusion from BI violation. Since signal locality is an empirically testable (and well-tested) consequence of relativity, violation of BI would imply unpredictability. This understanding actually plays the fundamental role in an important information theoretic protocol, namely, device independent randomness certification, a pioneer work done by Pironio et al. \cite{rand1}. Before describing the work of Pironio \emph{et al.} we first briefly discuss the framework of device independent scenario.

\subsection{Device independent scenario}
The necessity of device independent technique in quantum information processing tasks was pointed out by A. Ac\'{\i}n \emph{et al.} \cite{Masanes}. They observed that in the BB84 quantum key distribution (QKD) protocol \cite{BB84} the security proof was based on
\begin{itemize}
\item[(i)] the validity of the quantum formalism,
\item[(ii)] the legitimate partners perfectly know how their correlation is established, e.g., they know the dimensions of the Hilbert space describing their quantum systems.
\end{itemize}
The security proofs of QKD schemes exploit the well-established quantum features such as the no-cloning theorem, or the monogamy (i.e., nonshareability) of of certain quantum correlations. But if there is no restriction on the Hilbert space dimension, the security proof breaks down. In the noise-free case, the BB84 correlation satisfies $p(a=b|x=y)=1$ and $p(a=b|x=y)=1/2$. If such correlation results from measurements in the $x$ and $z$ bases on qubit pairs, then the state of these two qubits is necessarily maximally entangled and security follows from monogamy of entanglement. However, the authors, in \cite{Barrett}, pointed out that the same correlation can also be reproduced by the following four-qubit state:
\begin{equation}
\rho_{AB}=\frac{1}{4}(|00\rangle\langle 00|_z+|11\rangle\langle 11|_z)\otimes(|00\rangle\langle 00|_x+|11\rangle\langle 11|_x)
\end{equation}  
Here, Alice holds the first and third qubit. Whenever she measures in the $z$ ($x$) basis, she is actually measuring the first (third) qubit in this basis. The same happens for Bob, with the second and fourth qubit. It is easy to verify that the correlation is precisely same as ideal BB84 case. As the state is separable, so a secure key cannot be established. Thus key distribution in these schemes is successful when assumption (ii) is satisfied.

\begin{figure}[t!]
\centering
\includegraphics[height=5cm,width=7cm]{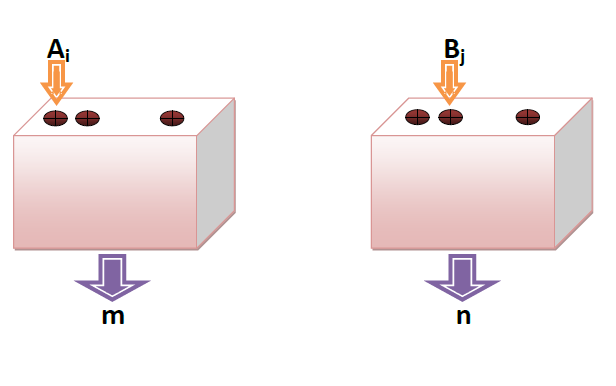}
\caption{Black-box description of an experimental set up used in device independent protocol.}\label{pic1}
\end{figure}

In device independent scenario one  does not have detail knowledge about the experimental apparatus, i.e., the experimental set up is like a black-box as shown in Fig(\ref{pic1}). On each side, the experimental device is depicted like a box with some knobs. A knob with different positions on each device, denoted respectively by $A_i$ and $B_j$, allows Alice and Bob to change the parameters of each measuring apparatus.  
Each measurement performed by Alice and Bob has $d$ possible outcomes. Finally, the frequencies $P(A_i=m,B_j=n)$ of occurrence of a given pair of outcomes for each pair of measurements have to be collected. Then it has to be checked whether the correlation satisfy some established physical principles. Only if it fulfills the required conditions, the correlation can be used for secure protocols. In the following we discuss such a device independent task, namely,  device independent randomness certification.   

\subsection{DI randomness certification from BI}
In DI randomness certification scenario one (Say Alice) has a private place which is completely inaccessible from outside i.e., no illegitimate system may enter in this place. From a cryptographic point of view assumption of such private place is admissible. Alice  choses classical inputs $x\in X$ and $y\in Y$ with probability distributions $\mathcal{P}_X(x)$ and $\mathcal{P}_Y(y)$, respectively, and sends them to two measurement devices ($\mathcal{MD}1$ and $\mathcal{MD}2$ respectively) through some secure classical communication channels. The inputs prescribe the  measurement devices to perform some POVM $\{M_{a|x}~|~M_{a|x}\ge 0~\forall a;~\sum_{a}M_{a|x}=\mathbb{I}_{\mathcal{H}_A}\}$ and $\{M_{b|y}~|~M_{b|y}\ge 0~\forall b;~\sum_{b}M_{b|y}=\mathbb{I}_{\mathcal{H}_B}\}$ on some quantum state $\rho$, shared between the two devices. Once the inputs are received, no classical communication between the measurement devices $\mathcal{MD}1$ and $\mathcal{MD}2$ is allowed. Alice collects the input-output statistics $P(AB|XY)=\{p(ab|xy)\}$. 
\begin{figure}[b]
\centering
\includegraphics[height=7cm,width=9cm]{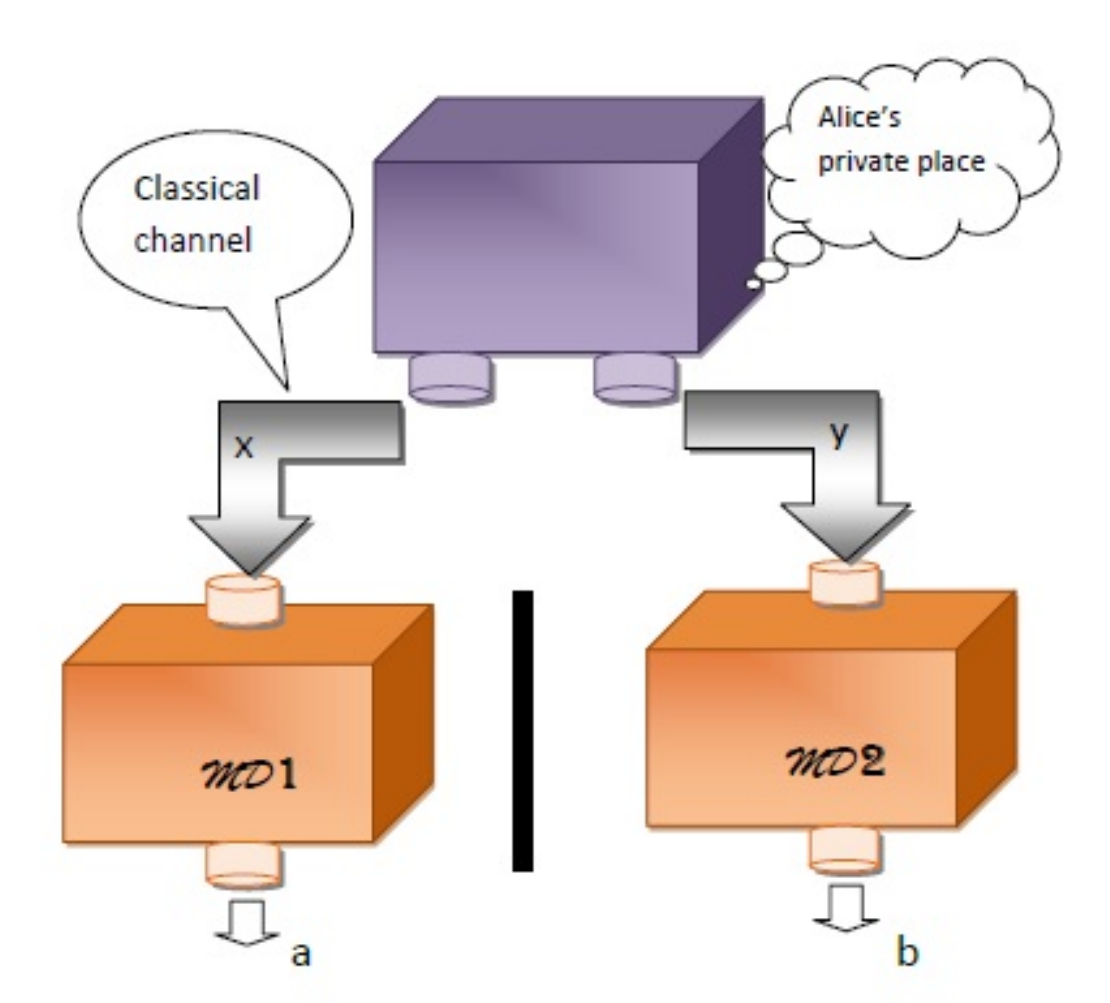}
\caption{Setup for DI randomness certification. Classical inputs are sent from Alice's private place to the measurement devices ($\mathcal{MD}1$ and $\mathcal{MD}2$) through secure classical channels. Classical communication is not allowed between two measurement devices.}\label{fig1}
\end{figure}
If the input-output statistics violate the Bell expression $I$, then min-entropy associated with the output is non zero and thus certifiable randomness is obtained. In this case the randomness is certified by the Bell's theorem. As the allowed correlations satisfy signal locality, BI violation implies that operational statistic must be unpredictable and thus randomness can be obtained without knowing the detail of the experimental device. The setup for DI randomness certification is depicted in Fig.\ref{fig1}. To obtain a lower bound in the min-entropy corresponding to a given Bell violation $I$ which does not depend on the input pair one has to first optimize the guessing probability over the all input pair $(x,y)$. If this optimized value is denoted as $G^*$ then $-\log_2G^*$ denote the minimum amount of random bits associated with Bell violation $I$. 

Let Alice is interested in the amount of minimum randomness obtained in NS theory; which mean that any correlation satisfying NS condition is allowed to share between the measurement devices. In that case, the following optimization problem need to be solved:    
\begin{eqnarray}\label{rand_ns}
p^*_{ns}(ab|xy)&=&~~~~~~\mbox{max}~~~~~~p(ab|xy)\nonumber\\
&&\mbox{subject~to}~~\sum_{abxy}c_{abxy}p(ab|xy)=I\nonumber\\
&& P(AB|XY)\in \mathcal{P}^{NS}. 
\end{eqnarray} 
An input-output probability distribution would lie in $\mathcal{P}^{NS}$ if the following conditions are satisfied: 
\begin{itemize}
\item[(a)] $p(ab|xy)\ge 0$ $\forall~~a,b,x,y$ 
\item[(b)] $\sum_{a,b}p(ab|xy)=1$ $\forall~~x,y$ 
\item[(c)] $\sum_{b}p(ab|xy)=\sum_{b}p(ab|xy')$ $\forall~~x,y,y'$
\item[(d)] $\sum_{a}p(ab|xy)=\sum_{a}p(ab|x'y)$ $\forall~~x,x',y$
\end{itemize}
where the first and second conditions, respectively, denote positivity and normalization of a probability distribution, the last two conditions are known as NS condition. The minimum amount of random bits corresponding to Bell violation $I$ in NS framework is therefore $H_\infty(AB|XY)=-\log_2\max_{ab}p_{ns}^*(ab|xy)$.

To obtain the minimum randomness in quantum theory one has to perform the following optimization problem: 
\begin{eqnarray}\label{rand_quantum}
p^*_q(ab|xy)&=&~~~~~~\mbox{max}~~~~~~p(ab|xy)\nonumber\\
&&\mbox{subject~to}~~\sum_{abxy}c_{abxy}p(ab|xy)=I\nonumber\\
&&p(ab|xy)=\mbox{tr}[M_{a|x}\otimes M_{b|y} \rho]
\end{eqnarray}
where the optimization is performed over all states and all POVMs defined over Hilbert space of arbitrary dimension. The second condition is to ensure that the obtained correlation is quantum one. Adapting a straightforward way of technique introduced in \cite{Navascues}, one can efficiently check whether a given correlation can be obtained via quantum means or not. The minimum random bits obtained in quantum theory corresponding to BI violation $I$ is thus quantified as $H_\infty(AB|XY)=-\log_2\max_{ab}p_q^*(ab|xy)$. From the analysis of Ref.\cite{rand1} it turns out that the minimum amount of randomness corresponding to a Bell violation $I$ obtained from quantum mechanics is higher than that obtained under consideration of NS framework.

Nonlocal correlation finds applications in important information theoretic tasks and hence is considered as useful resource. Quantifying the amount of nonlocality in a given correlation is, therefore, important from practical point of view. The amount of BI violation, beyond the \emph{local-realistic} bound, gives a natural measure of nonlocality.  This measure also has a operational interpretation as it gives the success probability of winning a nonlocal game (eg. for the two-input two-output scenario the well known XOR game). Though the optimal success probability in QM is strictly greater than classical theory, it must be restricted in comparison to more general correlations compatible with relativistic causality. In the following we discuss this nontrivial aspect of quantum nonlocal correlations.  

\section{Nonlocality in quantum theory: Cirel'son bound}\label{sec5}
Popescu and Rohrlich first gave an example of a correlation which is more nonlocal than quantum correlations but still satisfy the relativistic causality or no signaling (NS) principle. There correlation reads as:
\begin{eqnarray}\label{PR}
p(ab|xy)&=&\frac{1}{2}~~\mbox{if}~~a\oplus b=xy,\nonumber\\
&=& 0~~\mbox{otherwise}.
\end{eqnarray}

While in a generalized non-signaling theory Bell expression (left hand side of Eq.(\ref{BI})) can reach the maximum algebraic value $4$ i(t is easy to verify that the correlation of Eq.(\ref{PR}) achieves the maximal value $4$) in QM this value is restricted by $2\sqrt{2}$. We have already seen that performing suitable dichotomic measurements on singlet state the value $2\sqrt{2}$ can be achieved. However, Cirel'son showed that sharing any bipartite state and performing any pairs of dichotomic observables on each part the maximum value of Bell expression can not go beyond $2\sqrt{2}$ \cite{Cirelson}. This value is known as \emph{Cirel'son bound}.  

So, on the one hand QM contains astonishing nonlocal correlation, on the other hand, surprisingly, the nonlocality in QM is restricted is some sense. In the recent past researchers have tried to find out physical principles that can explain this restricted nonlocal feature of QM. Various interesting results have been derived in this regard. First successful attempt was made by W. van Dam \cite{van-Dam}. The result is further generalized by Brassard \emph{et al.}, who showed that in any world in which communication complexity is nontrivial, there is a bound on how much nature can be nonlocal \cite{Brassard}. Pawlowski \emph{et al.} introduced a new information principle, namely Information Causality (IC), which says that information that Bob can gain about a data set with Alice, by using all his local resources (which may be correlated with Alice's resources) and a classical communication from her, is bounded by the information volume of the communication \cite{Pawlowski}. The authors further showed that for any theory, satisfying IC, the Bell quantity can not be larger than $2\sqrt{2}$. on the other hand, Navascues \emph{et al.} postulated that any post-quantum theory should recover classical physics in the macroscopic limit \cite{Navascues}. Using this mechanism they were able to derive nontrivial bound on the strength of correlations between distant observers in any physical theory. Recently, another principle, namely, Local orthogonality \cite{Fritz} or Exclusivity \cite{Cabello1,Amaral}, an intrinsically multi-partite principle, have been used for the same purpose. Furthermore it has been shown that this principle also exactly singles out the Cirel'son bound \cite{Cabello2}. Recently, there is a different kind of development where Uncertainty principle and Complementarity principle have been shown to be related to the nonlocality of a theory in a fundamental way. In the following we discuss these two aspects \cite{Oppenheim,Banik}.

\subsection{Uncertainty, Steering and Nonlocality} 
In a recent paper Oppenheim and Wehener have shown that QM cannot be more nonlocal with measurements that respect the uncertainty principle \cite{Oppenheim}. In fact they have proved that degree of nonlocality of any theory is determined by two factors: the strength of the uncertainty principle and the strength of a property called ``steering''. Important to note that they have used a new mathematical form, namely the \emph{fine grained uncertainty}, to establish their result. 

{\bf Fine grained uncertainty}: Heisenberg’s uncertainty principle tells that there are incompatible measurements whose results cannot be simultaneously predicted with certainty \cite{Heisenberg}. Mathematically, Heisenberg expressed this fact in term of a nonzero lower bound on the product of the standard deviation of the concerned incompatible measurements. There are more modern approach to express this impossibility in term of entropic measures \cite{Entropic}. Oppenheim \emph{el al.} realized that entropic functions are, however, a rather coarse way of measuring the uncertainty and they have introduced the `fine grained' form of the uncertainty principle. Let $p(x^{(t)}|t)_{\sigma}$ denote the probability that we obtain outcome $x^{(t)}$ when performing a measurement labeled $t\in\mathcal{T}$ when the system is prepared in the state $\sigma$ and let $\mathbf{x}=(x^{(1)},x^{(2)},...,x^{(n)})\in\mathcal{B}^{\times n}$, with $n=|\mathcal{T}|$, denote the combination of possible outcomes. Then for a fixed set of measurements and a probability distribution $\mathcal{D}=\{p(t)\}_t$ the set of inequalities 
\begin{equation}\label{fgu}
U=\left\lbrace \sum_{t=1}^{n}p(t)p(x^{(t)}|t)_{\sigma}\le\zeta_{\mathbf{x}}~
|~\forall~\mathbf{x}\in\mathcal{B}^{\times n}\right\rbrace 
\end{equation}
describe  a fine-grained uncertainty relation. This relations tell that whenever $\zeta_{\mathbf{x}}<1$ one cannot obtain a measurement outcome with certainty for all measurements. The quantity
\begin{equation}
\zeta_\mathbf{x}=\max_{\sigma}\sum_{t=1}^{n}p(t)p(x^{(t)}|t)_{\sigma}
\end{equation}
characterizes the `amount of uncertainty' in a particular physical theory. Here, maximization is taken over all states allowed on a particular system. As for example, in quantum theory $\zeta_\mathbf{x}^Q=\frac{1}{2}(1+\frac{1}{\sqrt{2}})$, on the other hand in classical theory as well as in the box theory (marginal correlation for PR correlation or in other word for g-bit) $\zeta_\mathbf{x}^C=\zeta_\mathbf{x}^{box}=1$.

{\bf Steering}: Steering denotes the power of remotely preparing different ensembles of a system's state at one place from another spatially separated place. In quantum theory this concept 
was first introduced by Schr\"{o}dinger \cite{Schro}. We know that a mixed state $\sigma$ of a quantum system can be decomposed in many different ways as a convex sum:
\begin{equation}
\sigma=\sum_jp_j\sigma_j,
\end{equation}
where $\sigma_j$'s are density operators and $\{p_j\}_j$ denote probability distribution. This can be written as
an ensemble representation $\mathcal{E}=\{p_j,\sigma_j\}_a$ of the state $\sigma$. 
If Alice and Bob share a pure entangled state $\sigma_{AB}$ with the reduced state $\sigma_B=\mbox{tr}_A(\sigma_{AB})=\sigma$
on Bob's side then corresponding to every ensemble $\mathcal{E}$ there exists a measurement on Alice’s system that allows her to prepare the ensemble \cite{gisin,hjw}. It is important to note that this steering phenomena does not violate no-signaling 
principle as the unconditional state on Bob's side remains same for Alice's different measurements. As an precise example, consider that Alice and Bob share the singlet state $|\psi^-\rangle_{AB}$. Consider two different ensembles $\{\frac{1}{2},\frac{1}{2},|0_z\rangle\langle0_z|,|1_z\rangle\langle1_z|\}$ and $\{\frac{1}{2},\frac{1}{2},|0_x\rangle\langle0_x|,|1_x\rangle\langle1_x|\}$ for Bob's reduced state. Alice can prepare these two ensemble by performing measurement on her particle along $z$-direction and $x$-direction, respectively. However, it has been later observed that steering is not a strict quantum phenomenon, rather there exist broad class of theories which exhibit this \emph{non classical} feature \cite{Oppenheim,Barnum,Spekkens}.

With these two ideas, Oppenheim \emph{et al.} succeeded to establish that for any physical theory, the strength of nonlocal correlations is determined by a trade off between two aspects: steerability and uncertainty. In the following we discuss this trade off for some interesting class of theories. 
\begin{enumerate}
\item[(a)] \emph{Box world}: Here the power of steering is compatible to no signaling principle and there is no uncertainty in this theory. As a result the Bell expression in this theory obtained it's algebraic maximum value $4$.  
\item[(b)] \emph{Q. theory}: Here, also, the ability to steer is only limited by the no-signaling principle, i.e., maximal. But, due to presence of uncertainty the nonlocal power of this theory is restricted by Cirel'son bound.
\item[(c)] \emph{Classical theory}: In classical theory there is no restriction in knowing the values of different observables, simultaneously, or in other word we have no uncertainty relations on the full set of deterministic states. But, this theory shows no nonlocal property as steering is absent in this theory. 
\item[(d)] \emph{Local theory}: On the other hand there exists local hypothetical theories, which have perfect steering, but only due to a high degree of uncertainty they do not exhibit nonlocal behavior. Spekkens toy-bit model is one interesting example of such a theory \cite{Spekkens}. It has steering comparable to that in QM and Box world. Had there  been no uncertainty, nonlocality would have been equal to that of Box world. However, it can be shown that $\xi_{\mathbf{X}}^{toy}=\frac{1}{2}$ and hence the theory turns out to be local one.  
\end{enumerate}
It is important to note that there is no difference among the theories described in (c) and (d), so far their nonlocal behavior is concerned. However, the other properties of these theories are completely different.

\subsection{Complementarity and degree of nonlocality}
One of the original versions of the complementarity principle tells that there are observables in quantum mechanics that do not admit unambiguous joint measurement. Examples are position and momentum \cite{davice,prugo,bu}, spin measurement in different directions \cite{bu,kraus}, path and interference inn the double slit experiment \cite{scully,woot}, etc. With the introduction of the generalized measurement i.e. positive operator-valued measure (POVM), it was shown that observables which do not admit perfect joint measurement, may allow joint measurement if the measurements are made sufficiently fuzzy \cite{busch}. The optimal value of unsharpness that guarantees joint measurement of all possible pairs of dichotomic observables can be considered as the degree of complementarity and it has been shown that it determines the degree of nonlocality of the theory \cite{Banik}.

{\bf General framework}: Consider a generalized probability theory \cite{barrett} where any state of the system is described by an element $\omega$ of $\Omega$, the convex state-space of the system. $\Omega$ may be considered as a convex subset of a real vector space. By convexity of the state-space $\Omega$ we mean that any probabilistic mixture of any two states $\omega_1, \omega_2\in \Omega$, will describe a physical state of the system. An observable $\verb"A"$ (with the corresponding outcome set $\{\verb"A"_{j}~ : j \in J\}$) is an affine map from $\Omega$ into the set of probability distributions on the outcome set. A measurement of an observable $\verb"A" \equiv \{\verb"A"_j~|~\sum_jp^{\omega}_{\verb"A"_j}=1~~\forall~\omega\in\Omega\}$, performed on the system, allows us to gain information about the state $\omega$ of the physical system. The measurement of $\verb"A"$ consists of various outcomes $\verb"A"_{j}$ with $p^{\omega}_{\verb"A"_j}$ being the probability of getting outcome $\verb"A"_j$, given the state $\omega$. Let $\Gamma$ be the set of all observables with two measurement outcomes ( $j = + 1, - 1$ ), say `yes'$(=+1)$ and `no'$(=-1)$. If $\verb"A"\in\Gamma$ is such a kind of two-outcome observable, then the average value of $\verb"A"$ on a state $\omega$ is given by
\begin{equation}
\langle \verb"A"\rangle_\omega = p^\omega_{\verb"A"_{yes}}-p^\omega_{\verb"A"_{no}}.
\end{equation}
Given a two-outcome observable $\verb"A"\equiv\{\verb"A"_{yes},\verb"A"_{no}|~p^{\omega}_{\verb"A"_{yes}}
+p^{\omega}_{\verb"A"_{no}}=1~~\forall~\omega\in\Omega\}$, we define a fuzzy or unsharp observable, again with binary outcomes $\verb"A"^{(\lambda)}\equiv\{\verb"A"^{(\lambda)}_{yes},\verb"A"^{(\lambda)}_{no}~|~p^{\omega}_{\verb"A"^{(\lambda)}_{yes}}
+p^{\omega}_{\verb"A"^{(\lambda)}_{no}}=1~~\forall~\omega\in\Omega\}$, with `unsharpness parameter' $\lambda\in(0,1]$, where $p^{\omega}_{\verb"A"^{(\lambda)}_{yes(no)}}$ is the probability of getting the outcome $\verb"A"^{(\lambda)}_{yes(no)}$ in the measurement of $\verb"A"^{(\lambda)}$ with the result `yes' (`no'). The  probabilities $p^{\omega}_{\verb"A"^{(\lambda)}_{yes(no)}}$ are smooth versions of the probabilities of their original counterparts in the following way:
\begin{equation}
p^\omega_{\verb"A"^{(\lambda)}_{yes}}=\left(\frac{1+\lambda}{2}\right)p^\omega_{\verb"A"_{yes}}+\left(\frac{1-\lambda}{2}\right)p^\omega_{\verb"A"_{no}},
\end{equation}
for all $\omega\in\Omega$. We denote the set of all unsharp observables with binary outcomes for a given $\lambda$  by $\Gamma^{(\lambda)}$. For any $\verb"A"^{(\lambda)}\in\Gamma^{(\lambda)}$ the average value of $\verb"A"^{(\lambda)}$ on a given state $\omega\in\Omega$ can be calculated as:
\begin{equation}
\langle \verb"A"^{(\lambda)}\rangle_\omega = p^\omega_{\verb"A"^{(\lambda)}_{yes}}-p^\omega_{\verb"A"^{(\lambda)}_{no}}= \lambda \langle \verb"A"\rangle_\omega.
\end{equation}
Given a state $\omega\in\Omega$ and two observables $\verb"A"_1\equiv\{\verb"A"_{1j}~|~\sum_jp^{\omega}_{\verb"A"_{1j}}=1~~\forall~\omega\in\Omega\}$ and $\verb"A"_2\equiv\{\verb"A"_{2k}~|~\sum_kp^{\omega}_{\verb"A"_{2k}}=1~~\forall~\omega\in\Omega\}$, we say that joint measurement of $\verb"A"_1$ and $\verb"A"_2$ exists if there exists a joint probability distribution $\{p^{\omega}_{\verb"A"_{1j},\verb"A"_{2k}}|\sum_{j,k}p^{\omega}_{\verb"A"_{1j},\verb"A"_{2k}}=1\}$ satisfying the following conditions:
\begin{eqnarray}
\sum_{k}p^{\omega}_{\verb"A"_{1j},\verb"A"_{2k}}=p^{\omega}_{\verb"A"_{1j}},~~\forall~j,\nonumber\\
\sum_{j}p^{\omega}_{\verb"A"_{1j},\verb"A"_{2k}}=p^{\omega}_{\verb"A"_{1k}},~~\forall~k,
\end{eqnarray}
whatever be the choice of $\omega\in\Omega$. For our purpose, we will concentrate only on the existence of the joint measurement of two two-outcome observables $\verb"A"_1,\verb"A"_2\in\Gamma$.  Given a physical theory it is not justifiable to demand that joint measurement should exist for any pair of $\verb"A"_1,\verb"A"_2\in\Gamma$, although in the classical world, it is always possible to construct a joint measurement observable. On the other hand, there are certain observables in quantum theory which can not be jointly measured jointly.

{\bf Degree of complementarity}: It may be possible that observables which are not jointly measurable in a theory, may admit joint measurement for their unsharp counterparts within that theory. For two given observables, the values of unsharp parameter that make joint measurement possible, depend on the observables. Let $\lambda_{opt}$ denotes the optimum (maximum) value of the unsharp parameter $\lambda$ that guarantees the existence of joint measurement for $all$ possible pairs of dichotomic observables $\verb"A"^{(\lambda)}_1,\verb"A"^{(\lambda)}_2\in\Gamma^{(\lambda)}$. $\lambda_{opt}$ can then be considered as a property of that particular theory. It is obvious from the definition that joint measurement must exists for any two $\verb"A"^{(\lambda)}_1,\verb"A"^{(\lambda)}_2\in\Gamma^{(\lambda)}$, where $\lambda\leq\lambda_{opt}$. The value of $\lambda_{opt}$ measures the degree of complementarity of the theory in the sense that as $\lambda_{opt}$ decreases, the corresponding theory has more complementarity. Of course, finding the value of $\lambda_{opt}$ for a theory will depend on the details of the mathematical structure of the theory.

{\bf Bound on nonlocality}: Let us now consider the case of a composite system consisting of two subsystems with associated state spaces ${\Omega}_1$ and ${\Omega}_2$ respectively (in a no-signaling probabilistic theory). The state space of the composite system is defined to be ${\Omega}_1 \otimes {\Omega}_2$, which is again a convex subset of a real vector space, whereas, an observable $\verb"A"_{12}$ is an affine map (with outcome space $\{\verb"A"_{12}^{(j)}~ :~ j \in J\}$) from ${\Omega}_1 \otimes {\Omega}_2$ into the set of all probability distributions on the outcome space \cite{barrett}. Given any observable $\verb"A"_{1}$ for the first subsystem (with outcome space $\{\verb"A"_{1j} : j \in J_1\}$) and any observable $\verb"A"_{2}$ for the second subsystem (with outcome space $\{\verb"A"_{2k} : k \in J_2\}$), here, for our purpose, we consider only observables of the form $\verb"A"_{12} = \{\verb"A"_{12}^{(jk)} : p^{\eta}_{\verb"A"_{12}^{(jk)}} = p^{\eta}_{\verb"A"_{1j}, \verb"A"_{2k}}~ {\rm for}~ {\rm all}~ (j, k) \in J_1 \times J_2~ {\rm and}~ {\rm for}~ {\rm all}~ \eta \in {\Omega}_1 \otimes {\Omega}_2\}$ where $p^{\eta}_{\verb"A"_{1j}, \verb"A"_{2k}}$ is the probability of getting the result $(j, k)$ when measurement of $\verb"A"_{1}$ and $\verb"A"_{2}$ are performed on the joint state $\eta$. Thus, when we take the unsharp version $\verb"A"^{(\lambda)}$ of a dichotomic observable for the first subsystem and a dichotomic observable $\verb"B"$ for the second subsystem then, for any state $\eta \in {\Omega}_1 \otimes {\Omega}_2$, $p^{\eta}_{\verb"A"^{(\lambda)}_{yes}, \verb"B"_{yes}}$ will be the unsharp version of the probability $p^{\eta}_{\verb"A"_{yes}, \verb"B"_{yes}}$, {\it i.e.}, $p^{\eta}_{\verb"A"^{(\lambda)}_{yes}, \verb"B"_{yes}} = (1/2 + {\lambda}/2)p^{\eta}_{\verb"A"_{yes}, \verb"B"_{yes}} + (1/2 - {\lambda}/2)p^{\eta}_{\verb"A"_{no}, \verb"B"_{yes}}$, etc.

\textbf{Theorem 2}: Consider a composite system composed of two subsystem with state spaces $\Omega_1$ and $\Omega_2$ respectively in a no-signaling probabilistic theory. For any pair of dichotomic observabless $\verb"A"_1,\verb"A"_2\in\Gamma_1$  on the first system and dichotomic observables $\verb"B"_1,\verb"B"_2\in\Gamma_2$  on the second system with the joint state $\eta\in\Omega_1 \otimes \Omega_2$, we have the following inequality:
\begin{equation}\label{comple}
|\langle \verb"A"_1\verb"B"_1\rangle_\eta+\langle \verb"A"_1\verb"B"_2\rangle_\eta+\langle \verb"A"_2\verb"B"_1\rangle_\eta-\langle \verb"A"_2\verb"B"_2\rangle_\eta|\leq\frac{2}{\lambda_{opt}},
\end{equation}
where $\lambda_{opt}$ has the meaning as described above.

From the expression of inequality (\ref{comple}) it is clear that the amount of Bell violation is upper bounded by the unsharp parameter $\lambda_{opt}$, which is a characteristic of complementarity of that particular physical theory. As for example, in classical theory joint measurement of any two dichotomic observables is possible which means $\lambda_{opt}=1$. Contrary to this, we will show that, in quantum mechanics, the value of $\lambda_{opt}$ is $\frac{1}{\sqrt{2}}$ and this has been proved as a theorem in Ref\cite{Banik}.

\textbf{Theorem 3}: Given any $d$-dimensional quantum system, joint measurement for unsharp versions of any two dichotomic observables $\overline{\mathcal{M}}_1$ and $\overline{\mathcal{M}}_2$ of the system is possible with the largest allowed value of the unsharpness parameter $\lambda_{opt}=\frac{1}{\sqrt{2}}$.

\textbf{Outline of proof}: First we describe the condition of joint measurability of unsharp versions of two dichotomic projection valued measurements (PVM). Let $\mathcal{M}_j\equiv\{\wp_j=|\psi_j\rangle\langle\psi_j|,~ \mathcal{I}-\wp_j=|\psi^\bot_j\rangle\langle\psi^\bot_j|\}$ (for $j=1,2)$ be two dichotomic PVMs, where $|\psi_j\rangle$, $|\psi^\bot_j\rangle$ are normalized pure states such that $\langle\psi_j|\psi^\bot_j\rangle=0$. The unsharp version of $\mathcal{M}_j$ be denoted as $\mathcal{M}^{(\lambda)}_j\equiv \{\wp_j^{({\lambda})} \equiv 
\frac{(1 + {\lambda})}{2}{\wp_j} + \frac{(1 - {\lambda})}{2}({\mathcal{I}} - {\wp_j}), ({\mathcal{I}} - {\wp_j})^{({\lambda})} \equiv \frac{(1- {\lambda})}{2}{\wp_j} + \frac{(1 + {\lambda})}{2}({\mathcal{I}} - {\wp_j})\}$. Joint measurement of $\mathcal{M}^{(\lambda)}_j$'s is possible iff there there exists a POVM $\mathcal{M}^{(\lambda)}_{12}\equiv\{G_{++},G_{+-},G_{-+},G_{--}\}$ such that each $G_{ij}$ is a positive operator satisfying the following properties:
\begin{eqnarray}
G_{++}+G_{+-}+G_{-+}+G_{--}=\mathcal{I};\nonumber\\
G_{++}+G_{+-}=\wp^{(\lambda)}_1~;G_{-+}+G_{--}=(\mathcal{I}-\wp_1)^{(\lambda)};\nonumber\\
G_{++}+G_{-+}=\wp_2^{(\lambda)}~;~G_{+-}+G_{--}=(\mathcal{I}-\wp_2)^{(\lambda)}.
\end{eqnarray}
In the measurement of the POVM $\mathcal{M}^{(\lambda)}_{12}$ if $G_{++}$ `clicks', we say that both $\wp^{(\lambda)}_1$ as well as $\wp^{(\lambda)}_2$ have been `clicked', and so on.

P. Busch have shown that for the qubit system, the POVM $\mathcal{M}^{(\lambda)}_{12}$ will exist for all possible pairs of unsharp measurement iff $0< \lambda \leq \frac{1}{\sqrt{2}}$ \cite{busch}. Thus the maximum allowed value of the unsharpness parameter $\lambda_{opt}$ is $\frac{1}{\sqrt{2}}$. In the case of PVM, the factor $\frac{1+\lambda}{2}$ ($\frac{1-\lambda}{2}$) has been interpreted as degree of reality (unsharpness) \cite{busch}. Consider $d$ dimensional Hilbert space $\mathcal{H}$ and also consider two dichotomic PVMs, $\{P,\mathbf{1}-P\}$ and $\{Q,\mathbf{1}-Q\}$. There is an useful result in linear algebra, which guarantees the existence of an orthonormal basis such that $\mathcal{H}=\bigoplus_{\alpha=1}^k\mathcal{H}_{\alpha}$, with $\mathcal{H}_{\alpha}$ at most two dimensional \cite{Halmos,Masanes}. Therefore we have:
\begin{eqnarray}
P=\oplus_{\alpha=1}^kP^{\alpha};~~~\mathbf{1}-P=\oplus_{\alpha=1}^k(\mathbf{1}-P)^{\alpha};\nonumber\\
Q=\oplus_{\alpha=1}^kQ^{\alpha};~~~\mathbf{1}-Q=\oplus_{\alpha=1}^k(\mathbf{1}-Q)^{\alpha},
\end{eqnarray} 
where $P^{\alpha},(\mathbf{1}-P)^{\alpha},Q^{\alpha},(\mathbf{1}-Q)^{\alpha}$ are null, one dimensional or two dimensional projectors on $\mathcal{H}_{\alpha}$. Using this result we extended Busch's result for any pair of two outcomes observables and reached at the Theorem 3 (see \cite{Banik} for detail).

As an immediate consequence of Theorem 2 and Theorem 3, it follows that the amount of nonlocality of quantum theory respects the  Cirel'son bound. According to Theorem 2, in a generalized probability theory the optimal BI violation is restricted to $\frac{2}{\lambda_{opt}}$. However, Theorem 2 can not determine whether this value will be achieved (or not) in a particular theory. In a recent work, Stevens and Busch have provided sufficient criteria for achieving this bound \cite{Stevens}.

\textbf{Theorem 4} (Stevens and Busch): In any probabilistic model of a system A that supports uniform universal steering, the Cirel'son bound is given by the tight inequality that can be saturated 
$$\mathbb{B}\le \frac{2}{\lambda_{opt}}.$$

In the Ref.\cite{Stevens} the authors have provided a simple nonclassical, nonquantum example, namely, \emph{squit} or Box theory (which is the marginal state space of the PR correlation). The two-dimensional state space of squit is given by a square, denoted as $\square$, which contains all the points $(x,y,1)$ with $-1\le x+y\le 1$, $-1\le x-y\le 1$. The collection of effects on $\square$ is a convex set $\mathcal{E}(\square)$ in the ordered linear space $\mathcal{A}(\square)$ of affine functionals on $\square$, i.e.
$$\mathcal{E}(\square):=\{e\in\mathcal{A}(\square)|0\le e(\omega)\le 1, ~\forall~\omega\in\square\}.$$

The squit leads to maximally incompatible effects in the sense that it leads to the smallest possible value of $\lambda_{opt}^{squit}=\frac{1}{2}$. As the correlated system (PR correlation) satisfy the sufficient criterion stated in Theorem 4, hence this theory achieves the maximal BI violation $4$.

\section{Conclusion}
Discovery of Bell's theorem is sometime considered as the most profound discovery of physical science of the last Century. Apart from its implication concerning world views about properties of physical world, the violation of Bell's inequality by quantum theory finds real application in information theory and communication tasks. In this review, we present various derivation of Bell's inequality with the aim of understanding the various implications of the violation of this inequality. We also discuss the recent discoveries of some universal principles (in the sense of encompassing all physical theories) and discuss how they can reproduce bound on Bell's inequality violation which exactly matches with the quantum bound. Finally we discuss the recent development regarding the relation of nonlocality with uncertainty and complementarity principles where uncertainty along with steering capacity determines the amount of nonlocality and complementarity provides an upper bound on nonlocality of the theory.        

\section*{Acknowledgments}
We gratefully acknowledge discussions with Sibashis Ghosh and Md. Rajjak Gazi.

\end{document}